# Compressed Sensing Applied to Weather Radar

Kumar Vijay Mishra, Anton Kruger, and Witold F. Krajewski
The University of Iowa, Iowa City, USA
kumarvijay-mishra@uiowa.edu

*Abstract*—We propose an innovative meteorological radar, which uses reduced number of spatiotemporal samples without compromising the accuracy of target information. Our approach extends recent research on compressed sensing (CS) for radar remote sensing of hard point scatterers to volumetric targets. The previously published CS-based radar techniques are not applicable for sampling weather since the precipitation echoes lack sparsity in both range-time and Doppler domains. We propose an alternative approach by adopting the latest advances in matrix completion algorithms to demonstrate the sparse sensing of weather echoes. We use Iowa X-band Polarimetric (XPOL) radar data to test and illustrate our algorithms.

## I. INTRODUCTION

Over the last two decades, weather radar has proven to be a valuable instrument providing critical precipitation information through remote sensing of the atmosphere [1]. Modern weather radar systems require an accurate and efficient information abstraction of voluminous data in real time. For example, a typical pulsed Doppler weather radar system samples data at 1000 *range bins* at 1 kilohertz pulse repetition frequency (PRF), generating approximately a million samples of complex *time series* data per second. The volume of time series data is sufficiently large that, in general, its storage for later analysis is impractical and advanced real time signal processing is a challenge [2]. One of the essential functions of weather radar signal processing is to reduce the data volume by restyling time series data into more meaningful meteorological *data products*.

The volume of radar signal acquisition is greatly affected by the *dwell time*, i.e., the time duration a radar beam spends hitting a particular target [3, 4]. Surveillance radars constantly scan the *target scene* in order to provide quick updates on the movement of targets-of-interest. The scan rate of the radar is often limited by the dwell time, which, in turn, is determined by the precision necessary to ascertain information about the target. A longer dwell time would lead to a more accurate target information, but, simultaneously, it would decrease the update rate of the target scene. Using conventional radar hardware and signal processing techniques, it is difficult to maintain the data quality with faster scan rates.

Despite the initial massive time series data acquisition by the weather radar, the end products are often reasonably downsized in volume leading to a pertinent question: could lesser samples have been acquired in the first place? In this paper, we present a novel meteorological radar sampling and processing scheme that allows for smaller dwell times without significant loss of target information. Our philosophy is to adopt recent advances in *compressed sensing* (CS) for the particular application of weather radar. CS is a novel signal processing technique that unites sampling and digital data compression in a single step by relying on the inherent *sparsity* of the data in some *dictionary* [5]. While conventional signal processing methods sample at Nyquist-Shannon rate and then compress the data for minimal storage, CS allows sampling of only useful information at lower sampling rates.

### A. Relation to prior work

Much of the existing research on applying CS to radar applications is devoted to *hard target radars* [6], where the target is contained in a single range bin resulting in sparsity in both range-time and frequency. Another natural candidate for CS application is multiple-input-multiple-output (MIMO) radar system where, in addition to range and Doppler, the radar signals are also sparse in *angle space* [7, 8]. CS application to radar imaging as that of terrain using synthetic aperture radar (SAR), isolated targets using inverse synthetic aperture radar (ISAR), and through-the-wall targets using ultra-wideband radars is another major research focus [6, 9]. CS-based imaging [10] is shown to reduce receiver hardware complexity by eliminating pulse compression [6], decrease data sampling cost [11] and enhance target detection resolution [12].

Among *volumetric target radars*, [13] uses compressed sensing to achieve better noise removal and high resolution detection of meteors which, though distributed over many range cells, are assumed to be highly localized in both range-time and frequency domains. There have been some attempts to use CS in weather radars for specialized tasks such as refractivity retrieval [14] by using the sparsity of refractivity difference in the discrete cosine transform (DCT) domain and making measurements using a phased-array antenna [15]. A related meteorological application is downscaling [16], where the sparsity of the rainfall *image* in wavelet domain is used to form a high-resolution precipitation image from low-resolution measurements [17]. Downscaling adds details to a low-resolution image using CS, but the radar data are obtained using conventional scanning and long dwell times. Also, CS-based downscaling is applied on the image of the rainfall data rather than directly on received radar signal.

The potential of CS in radar remote sensing of precipitation targets is relatively unexamined so far. The CS techniques developed for the hard target radars cannot be directly applied to weather radars because the precipitation echoes may not necessarily have sparsity in either range-time or Doppler. We resolve this issue by modeling the remote sensing of weather as a low-rank *matrix completion* problem since weather radar returns exhibit high spatial and temporal correlations [2].

This research is supported by Iowa Flood Center.

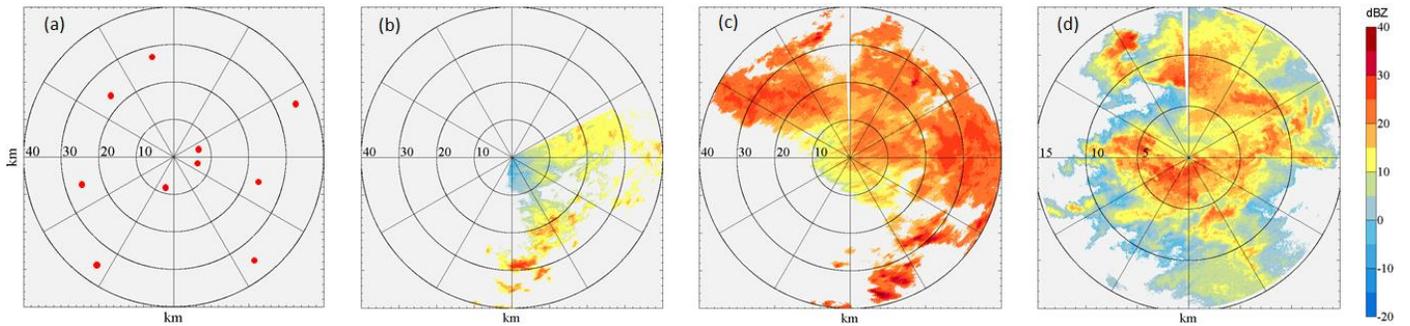
Fig. 1. (a) Point targets on a radar display. (b) Spatially sparse precipitation (Iowa XPOL-2 radar data observed on 0139 UTC, Jun 13, 2013). (c) Precipitation echoes are sparse along only a few range profiles (0150 UTC, Jun 13, 2013). (d) Precipitation returns do not exhibit spatial sparsity (2308 UTC, Jun 12, 2013).

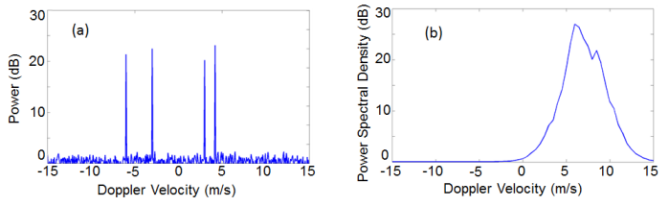
Fig. 2. Typical Doppler signatures for (a) point targets and (b) weather echoes in white noise using pulse repetition frequency = 2 kHz for an X-Band radar. Two incoming and two outgoing point targets are shown in (a). In (b), the estimated parameters are mean velocity ≈ 6 m/s and spectrum width ≈ 3.5 m/s.

*B. Our contributions*

Matrix completion approach for sparse sensing of point targets has been previously proposed for a colocated MIMO radar [7], even though sparsity already exists in range-Doppler-angle space. However, this kind of remodeling is not applicable to weather radar signals as they usually lack sparsity in such conventional domains. We propose sparse sampling of precipitation as partially observing a low-rank matrix and then view reconstruction of the target *scene* as recovering the full matrix. Unlike several other CS-based radar studies, we use real data from Iowa XPOL radars [18] to verify our algorithms.

## II. GENERAL CONSIDERATIONS

The compressed sensing deals with an NP-hard problem of minimizing the $\ell_0$ norm [5] i.e., the number of non-zero vector elements. The key result from the theory of compressed sensing states that, under certain "incoherency" conditions on the measurement process, this NP-hard problem can be replaced by its closest convex approximation: minimizing the $\ell_1$ norm. For CS techniques to succeed the radar signals must be sparse and measurements must be incoherent.

*A. Sparsity*

In the context of scanning radar, the spatial domain often has the same connotation as the time domain since every spatially distinct return is acquired at a different time instant. Let $s_{tx}(t)$ be the complex transmitted signal waveform at time $t$. Then, the received signal from the scatterer at range $r$ is given by,

$$s_{rx}(t) = a(r;t)s_{tx}(t - (2r/c)) \qquad (1)$$

where $a(r;t)$ is the time-varying scattering amplitude or target reflectivity and $c$ is the speed of light. The received voltage is the complex envelope of this received signal. If there are multiple scatterers along the range, then the received signal is the convolution of transmitted signal with the reflectivity profile $A(r;t)$,

$$s_{rx}(t) = \int_0^{R_{max}} A(r;t)s_{tx}(t - (2r/c))dr \qquad (2)$$

where $R_{max}$ is the maximum unambiguous range of the radar. If the transmit pulse-width is $\tau_0$, then the entire range profile can be divided into $N$ range bins of length $r_0 = c\tau_0/2$. Here, $N = \lfloor R_{max}/r_0 \rfloor$. The scattering amplitude or reflectivity can be expressed as a discrete sum of the reflectivities of $K$ individual scatterers each located at range $r_k$ as,

$$A(r;t) = \sum_{k=1}^{K} a(r_k;t)\delta(r - r_k) \qquad (3)$$

where $\delta(r)$ is the Kronecker delta function. For point target radars, $K \ll N$ implying that the radar signal is sparse in spatial or range-time domain. For example, Fig. 1(a) illustrates a typical radar display for aircraft surveillance radar, in which every red dot indicates an aircraft. On the other hand, weather radars sample precipitation echoes which extend over several range cells, often the entire radar coverage range, so that it is quite common to have $K \approx N$. The meteorological reflectivity product $Z_h$ that has units of dBZ is given by [1, p.82],

$$Z_h(r) = \overline{P}_{rx}(r) + C + 20\log_{10} r \qquad (4)$$

where $\overline{P}_{rx}$ is the mean received power (in dBm), C is the radar constant, and r is the range (in km). Fig. 1(b)-(d) show that the $Z_h$ for precipitation need not always have spatial sparsity.

The radar signals from point targets are also sparse in frequency domain [19] because the corresponding spectrum shows only a finite number of Doppler frequencies $f_d$ (as shown in the simulation of Fig. 2(a)). Assuming $v$ to be the radial Doppler velocity of the target and $\lambda$ the operational radar wavelength, we have $f_d = 2v/\lambda$, which appears as a spike in the spectrum sampled at frequency $f_s = PRF/2$. Due to the simple scaling relationship (for a constant $\lambda$), $v$ and $f_d$ are often used interchangeably. In the case of weather, the appropriate spectrum descriptor is power spectral density $S(v)$, since here the reflectivity $A(r;t)$ is also a random process (due to random locations of the scatterers). However, unlike point targets, the power spectral density of the weather echo closely follows a Gaussian shape [1, p. 136] and occupies a continuum of spectrum (thereby excluding any possible spectral sparsity):

$$S(v) = \frac{\overline{P}_{rx}}{\sigma_v\sqrt{2\pi}} \exp\left[-\frac{(v-\overline{v})^2}{2\sigma_v^2}\right] \quad (5)$$

where $\overline{v}$ is the mean velocity of precipitation and the standard deviation $\sigma_v$ is known as the spectrum width (Fig. 2(b)). Therefore, in general, precipitation is not sparse in frequency domain as well.

## B. Sampling Schemes

Most of the proposed strategies to ensure incoherence require changes in the waveform ("Alltop" sequence) [12], or using multiple radars on ground [8]. However, the most relevant scheme for our work is randomly sampling range bins in the radar coverage area using an array of antenna elements, as previously proposed for point target radars in [19]. If the antenna is not phased array but, say a parabolic dish, such a random sampling can be achieved by randomly selecting only a few time samples to pass through the receiver. This approach would require scanning the entire target scene (and therefore not reduce the dwell time), and then discard many samples randomly. An alternative can be setting the radar antenna to scan at a very high scan rate. Under normal scanning speeds, a weather radar produces meteorological estimates by averaging over several azimuths. Therefore, should the radar to scan at a very fast scan rate, it would then dwell less on each azimuth and would randomly "miss" other azimuths (which would have been otherwise dwelled by the radar at a normal scan rate). This scan strategy would then produce randomly sensed samples (fewer than a slower scan rate) for each range cell.

## III. SPARSE SAMPLING OF WEATHER ECHOES

Although weather radar signals cannot be modeled as sparse in conventional dictionaries such as time and frequency, the backscattered signal in a weather radar is coherent [2]. In other words, the motion among the precipitation scatterers is small compared to the radar wavelength, so their relative positions produce highly correlated echoes from sample to sample and scan to scan. This inherent redundancy in weather radar signals implies that the range-azimuth scan of precipitation echoes can be modeled as a low-rank matrix. Low-rank matrices are the multi-dimensional equivalents of one-dimensional sparse vectors. Given a matrix $M \in \mathbb{R}^{m \times n}$, its singular value decomposition (SVD) is given by, $M = USV^T, U \in \mathbb{R}^{m \times r}, V \in \mathbb{R}^{r \times n}$, and $S = \text{diag}(\sigma_1, \cdots, \sigma_r)$, where $\sigma_1 \geq \cdots \geq \sigma_r > 0$ are the unique singular values and $r \leq \min(m,n)$ is the rank of the matrix. For a low-rank matrix, most of the diagonal elements of $S$ are zero such that, $r \ll \min(m,n)$. The best $r'$-rank approximation $\widetilde{M}$ of the matrix $M$ is given by zeroing out the $r - r'$ smallest singular values so that, $\widetilde{M} = U\widetilde{S}V^T$ and $\widetilde{S} = \text{diag}(\sigma_1, \cdots, \sigma_{r'}, 0, \cdots, 0)$.

As an illustration, in Fig. 3 we plot the singular values of the spatially non-sparse 1930 (range gates) by 413 (azimuthal rays) $Z_h$ data matrix corresponding to the actual Iowa XPOL-2 radar data of Fig. 1(d). We observe that most of the singular values are very small or close to zero (due to highly correlated spatial samples of weather backscatter). In Fig. 4, we show the effect of low-rank approximation on the original weather echo. It should be noted that a very small fraction of detailed features are lost in the low-rank approximations of weather radar reflectivity product, illustrating the *rank sparsity* in the real weather radar echoes. This makes it possible to apply current research in matrix completion to fully recover a low-rank matrix from a randomly observed sample of its entries [20].

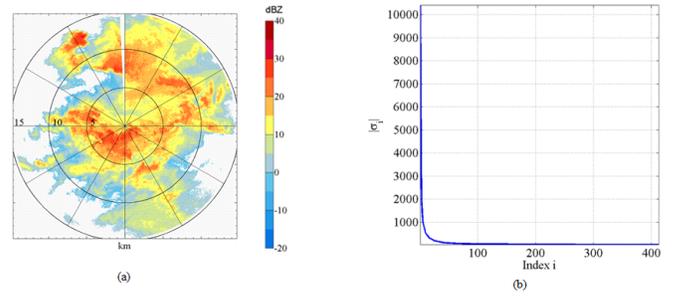

Fig. 3. (a) Original $Z_h$ data and (b) plot of its ordered singular values.

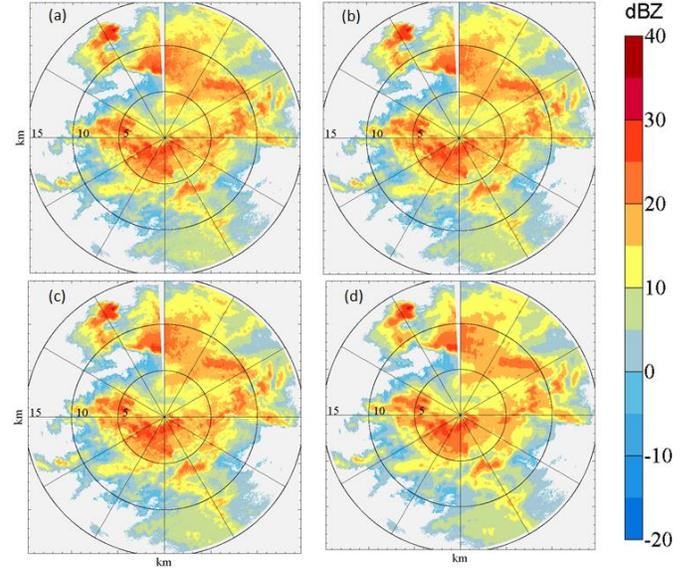

Fig. 4. (a) Full rank original $Z_h$. Low rank approximations of original data using (b) 40% (c) 25% (d) 5% of the most significant singular values.

Let $\Omega$ denote the set of the random locations of the partially observed entries of the original low rank matrix $M$. Then, recovering $M$ corresponds to the rank minimization problem:

$(P_0)$ $\quad$ minimize $rank(X)$
$\quad\quad$ subject to $X_{ij} = M_{ij}, (i,j) \in \Omega$ $\quad$ (6)

However, like $\ell_0$ minimization, rank minimization is also intractable. The approach to low-rank matrix completion is therefore to minimize the matrix equivalent of $\ell_1$, i.e., the nuclear-norm $\|X\|_* = \sum_k \sigma_k$,

$(P_1)$ $\quad$ minimize $\|X\|_*$
$\quad\quad$ subject to $X_{ij} = M_{ij}, (i,j) \in \Omega$ $\quad$ (7)

In Fig. 5, we reconstruct the sparsely sampled meteorological reflectivity using singular value thresholding (SVT) [21] for nuclear norm minimization. Although the illustrated reconstruction uses a low-rank approximation $\widetilde{Z}_h$ as matrix $M$, the results are not very different if $Z_h$ itself is used with some modifications to problem $P_1$ (the relative errors $\varepsilon_1$ and $\varepsilon_2$ are of the same order). The very close similarity of the reconstructed data distribution with the original clearly illustrates the potential of CS for weather radars.

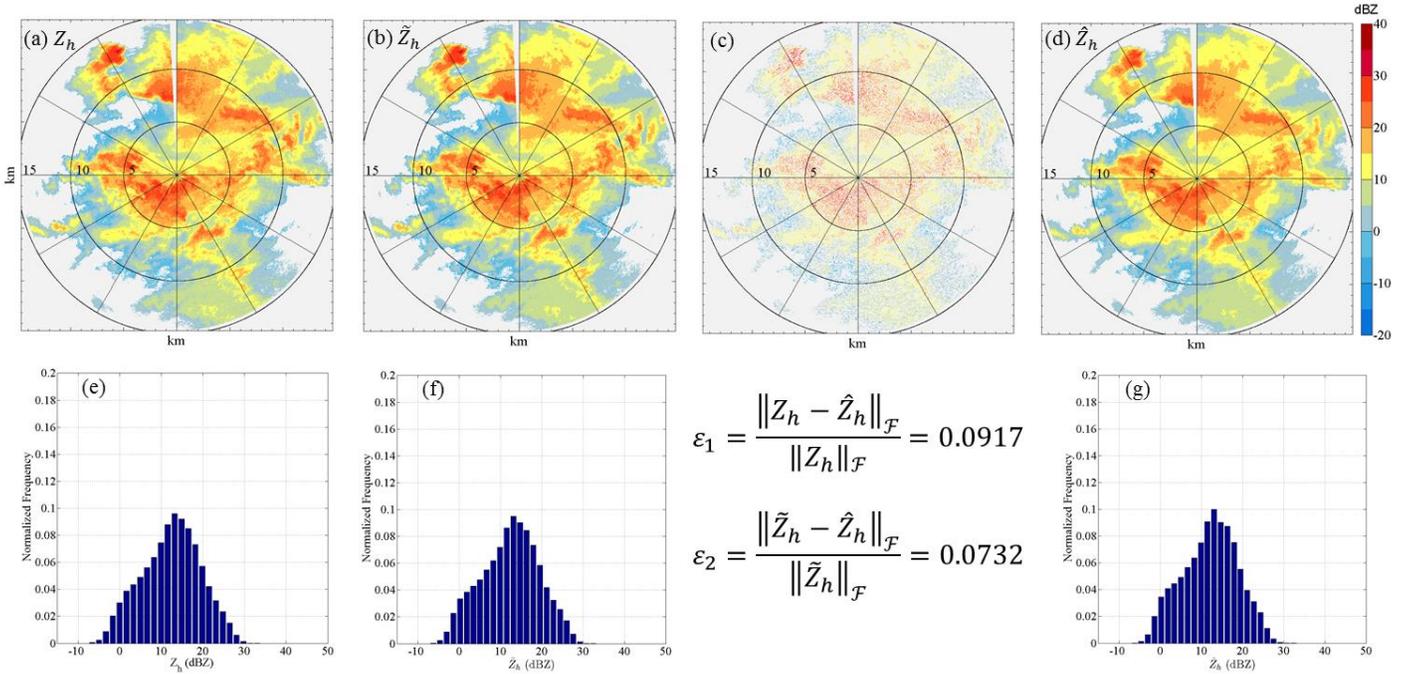

Fig. 5. (a) Original $Z_h$ (b) Low rank approximation $\tilde{Z}_h$ of original $Z_h$ obtained by retaining 25% of the most significant singular values (c) Randomly sampled one-third entries of $\tilde{Z}_h$ (d) Reconstructed reflectivity $\hat{Z}_h$ using SVT. Histograms of data corresponding to (e) $Z_h$ (f) $\tilde{Z}_h$, and (g) $\hat{Z}_h$. Two relative error metrics $\varepsilon_1$ and $\varepsilon_2$ are computed. $\|\cdot\|_{\mathcal{F}}$ denotes the Frobenius norm.

## IV. DISCUSSION AND SUMMARY

We proposed an unconventional weather radar paradigm that employs compressed sensing techniques to reduce the radar scan time without any significant loss of target information. We posed the sparse sampling of weather radar targets as a low-rank matrix completion problem, and verified our approach using real data from the Iowa XPOL radars. In this preliminary formulation of CS-based meteorological radar, we greatly simplified the radar signal model by ignoring the effects of receiver noise, ground clutter, mixed-phase hydrometeors and use of dual-polarization. In future work, it will be interesting to investigate these effects for the proposed CS-based meteorological radar.